\documentclass[amsmath,amssymb,aps,prl,twocolumn,superscriptaddress,showpacs,floatfix]{revtex4-1}

\usepackage{graphicx}
\usepackage[dvips]{color}
\usepackage{epstopdf}
\usepackage{ulem}

\newcommand\colorsout[1]{\bgroup \markoverwith{\textcolor{#1}{\rule[0.5ex]{2pt}{0.4pt}}}\ULon}

\usepackage{color}
\usepackage{braket}

\begin{document}

\title{Spin-flip scattering selection in a controlled molecular junction}
\author{M. Ormaza}
\email{ormaza@ipcms.unistra.fr}
\affiliation{Universit\'{e} de Strasbourg, CNRS, IPCMS, UMR 7504, F-67000 Strasbourg, France}
\author{P. Abufager}
\affiliation{Instituto de F\'{i}sica de Rosario, Consejo Nacional de Investigaciones Cient\'{i}ficas y T\'ecnicas (CONICET) and Universidad Nacional de Rosario, Bv. 27 de Febrero 210bis (2000) Rosario, Argentina}
\author{B. Verlhac}
\author{N. Bachellier}
\affiliation{Universit\'{e} de Strasbourg, CNRS, IPCMS, UMR 7504, F-67000 Strasbourg, France}
\author{M.-L. Bocquet}
\affiliation{PASTEUR, D\'epartement de Chimie, Ecole Normale Sup\'erieure, PSL Research University, Sorbonne Universit\'es, 
UPMC Univ. Paris 06, CNRS, 75005 Paris, France}
\author{N. Lorente}
\affiliation{Centro de F{\'{\i}}sica de Materiales CFM/MPC (CSIC-UPV/EHU), Paseo Manuel de Lardizabal 5, 20018 Donostia-San Sebasti\'an, Spain}
\affiliation{Donostia International Physics Center (DIPC), Paseo Manuel de Lardizabal 4, 20018 Donostia-San Sebasti\'an, Spain}
\author{L. Limot}
\email{limot@ipcms.unistra.fr}
\affiliation{Universit\'{e} de Strasbourg, CNRS, IPCMS, UMR 7504, F-67000 Strasbourg, France}
\date{\today}

\begin{abstract}
A simple double-decker molecule with magnetic anisotropy, nickelocene, is attached to the metallic tip of a low-temperature scanning tunneling microscope. In the presence of a Cu(100) surface, the conductance around the Fermi energy is governed by spin-flip scattering, the nature of which is determined by the tunneling barrier thickness. The molecular tip exhibits inelastic spin-flip scattering in the tunneling regime, while in the contact regime a Kondo ground state is stabilized causing an order of magnitude change in the zero-bias conductance. First principle calculations show that nickelocene reversibly switches from a spin $1$ to $1/2$ between the two transport regimes. 
\end{abstract}

\maketitle 


The magnetic anisotropy of surface-supported objects \textemdash atoms or molecules\textemdash has been attracting a growing interest in relation to ultra-dense storage technology~\cite{Gambardella2003,Donati2016} and quantum computing~\cite{Loss2001,molecularnanomagnets}. From a practical viewpoint, a uniaxial magnetic anisotropy energy of the form $D S_z^2$ provides stable magnetic states into which information can be encoded and may, moreover, be externally controlled~\cite{Hirjibehedin2006,Zyazin2010,Parks2010,Bryant2013,Oberg2014,Heinrich2015,Dubout15,Khajetoorians15,Jacobson2015,Ormaza2017}. Long-lived magnetic states are obviously desirable for data storage, but are keen to decoherence in surface-supported objects due to the effect of the environment~\cite{Gauyacq2015, Delgado2015, Delgado2017}. One of the most common effects, which involves the spin-flip elastic scattering of conduction electrons, is the Kondo screening occurring below a critical temperature $T_\text{K}$~\cite{Hewson97}. The general picture to address is that single objects have the potential for both magnetic anisotropy and the Kondo effect~\cite{Otte2008}, the outcome of their interplay depending on the object's spin~\cite{Zitko2009,Hurley2011,Korytar2012,Ternes2015} and on the relative weight of $k_\text{B}T_\text{K}$ versus $D$~\cite{Zitko2008,Misiorny2012}. 

For several years now scanning tunneling microscopy (STM) has proven to be well suited for exploring the spin-flip scattering occurring in surface-supported single objects~\cite{Gauyacq2012,Ternes2015}. Elastic and inelastic tunneling spectra can be used to probe the above mentioned interplay in an environment that can be characterized with atomic-scale precision. In recent experiments~\cite{Oberg2014,Khajetoorians15,Jacobson2015,Dubout15}, different surface topologies were used to tune the Kondo exchange interaction between the spin and the underlying metal, while spin control was possible through hydrogen doping of transition metal atoms. Here, we show that spin control can be achieved in a molecular junction using a functionalized STM tip. Up to now organic molecules have been used in scanning probe techniques to functionalize tips with the purpose of improving image resolution~\cite{Gross2009,Weiss2010,Schull2011, Gross2011} and, eventually, spectroscopic sensitivity~\cite{Ho2014, Guo2016}. In this work, we attach a $S=1$ molecule with an easy-axis magnetic anisotropy ($D>0$), nickelocene [Ni(C$_5$H$_5$)$_2$, noted Nc hereafter], to the tip apex of a STM and form a precise contact with a Cu(100) surface. We show that by moving Nc in and out of contact, we can control its spin and, concomitantly, the nature of the spin-flip scattering that governs the conductance around the Fermi energy of the junction. Our findings are corroborated by density functional theory (DFT) calculations.

\begin{figure}
  \begin{center}
    \includegraphics[width=0.47\textwidth]{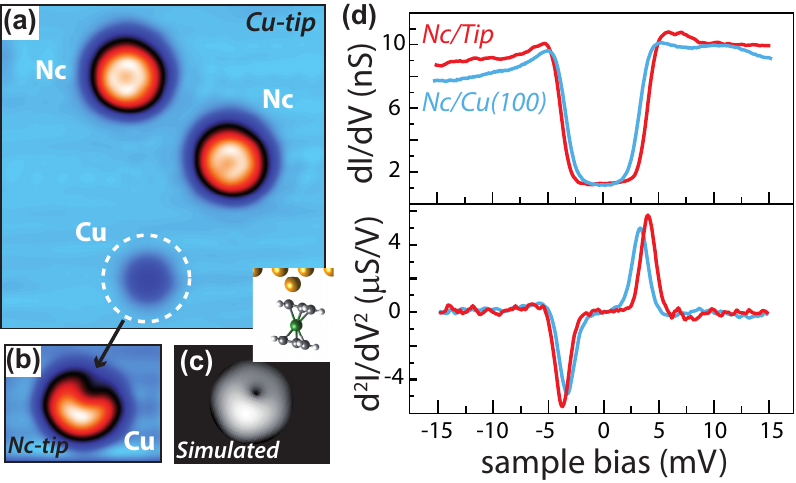}
  \end{center}
  \caption{(a) STM image acquired with a metallic tip ($5.3\times5.3$ nm$^2$, $-15$\,mV, $20$\,pA). (b) Counter image of the Cu atom obtained with a Nc-terminated tip ($1.7\times1.3$ nm$^2$, $-15$\,mV, $20$\,pA). (c) Calculated relaxed configuration of a Nc molecule on top of a Cu adatom on Cu(100), together with the simulated STM image. Atom colors: Cu (yellow), C (grey), H (white) and Ni (green). (d) $dI/dV$ spectra and their derivative acquired with a Cu-tip over a Nc molecule (solid blue line) and with a Nc-terminated tip over the Cu(100) surface (solid red line). Feedback loop opened at 100\,pA and -15\,mV.}
\label{fig1}
\end{figure}

The experiments were performed in an ultra-high vacuum STM operated at $2.5$~K. A clean Cu(100) surface was prepared by repeated cycles of Ar$^+$ sputtering and annealing up to $800$~K at a base pressure of $2\times10^{-10}$~mbar. Nickelocene molecules were deposited from a crucible onto the cold copper surface (below 80~K) at a rate of $0.025$\,ML/min resulting in a sub-monolayer coverage. Directly after molecular deposition, the sample was inserted in the STM. An image of the Cu(100) surface after a molecular deposition is shown in Fig.~\ref{fig1}(a), in which isolated Nc decorate the surface terraces. The molecules show a ring-shaped pattern, indicating that one cyclopentadienyl ring (C$_5$H$_5$, noted Cp hereafter) is exposed to vacuum, \textit{i.e.}, the long molecular axis of Nc is perpendicular to the Cu(100) surface~\cite{Bachellier2016}.


We found that in order to contact nickelocene between both electrodes, surface and tip, it is more stable to first transfer the molecule from the metallic surface to the tip \textemdash details of the molecule transfer to the metallic tip can be found in~\cite{Ormaza2017}. Prior to molecular transfer, care was taken to select a monoatomically sharp tip apex~\cite{Limot2005}. Information may be gathered on the status of Nc at the tip apex by acquiring counter-images ~\cite{Heinrich2011}, which consist in imaging single atoms on the surface with the molecular tip. The typical molecular pattern observed in the counter images of Cu adatoms is presented in Fig.~\ref{fig1}(b) and differs from the feature-less protrusion observed with a metallic tip [encircled in Fig.~\ref{fig1}(a)]. The pattern reveals that the tip is terminated by a Cp ring of a tilted Nc molecule, the tilt angle exhibiting some tip dependency. To confirm this assignment, we mimicked the molecular tip through DFT calculations by considering a Nc molecule adsorbed on a Cu atom on Cu(100) [inset of Fig.~\ref{fig1}(c), see Supplementary Material]. The molecule is undeformed, tilted by $13^{\circ}$ with respect to the surface normal and linked through two C atoms to the Cu atom~\cite{n1}. The corresponding simulated image in Fig.~\ref{fig1}(c) is in good agreement with the experimental counter-image of Fig.~\ref{fig1}(b).


The top panel of Fig.~\ref{fig1}(d) presents the $dI/dV$ spectrum acquired with a Nc-tip above Cu(100). For comparison, we have included the $dI/dV$ spectrum acquired with a metallic tip above a Nc molecule on Cu(100). Both spectra were acquired with a lock-in amplifier using a frequency of 716~Hz and an amplitude of 150~$\mu$V rms; the metallic tip was verified to have a flat electronic structure in the bias range presented. The steps, symmetric with respect to zero bias, that are visible in the $dI/dV$ correspond to the manifestation of efficient spin-flip excitations within the molecule~\cite{Ormaza2017}. These occur between the ground state $\ket{S=1, M=0}$ and the doubly degenerate $\ket{S=1, M=\pm1}$ excited states of the molecule, the threshold energy observed corresponding to the longitudinal magnetic anisotropy energy, $D$. The bottom panel shows the numerical derivative of the $dI/dV$ spectra in order to facilitate the identification of the threshold values. For Nc on the surface we find $D=(3.2\pm0.1)$~meV, while for Nc on the tip the magnetic anisotropy grows to $D=(3.7\pm0.3)$~meV. This reflects changes in the local ligand field induced by the different adsorption configuration~\cite{Bryant2013,Heinrich2015,Ormaza2017}. 

\begin{figure}
  \begin{center}
    \includegraphics[width=0.47\textwidth]{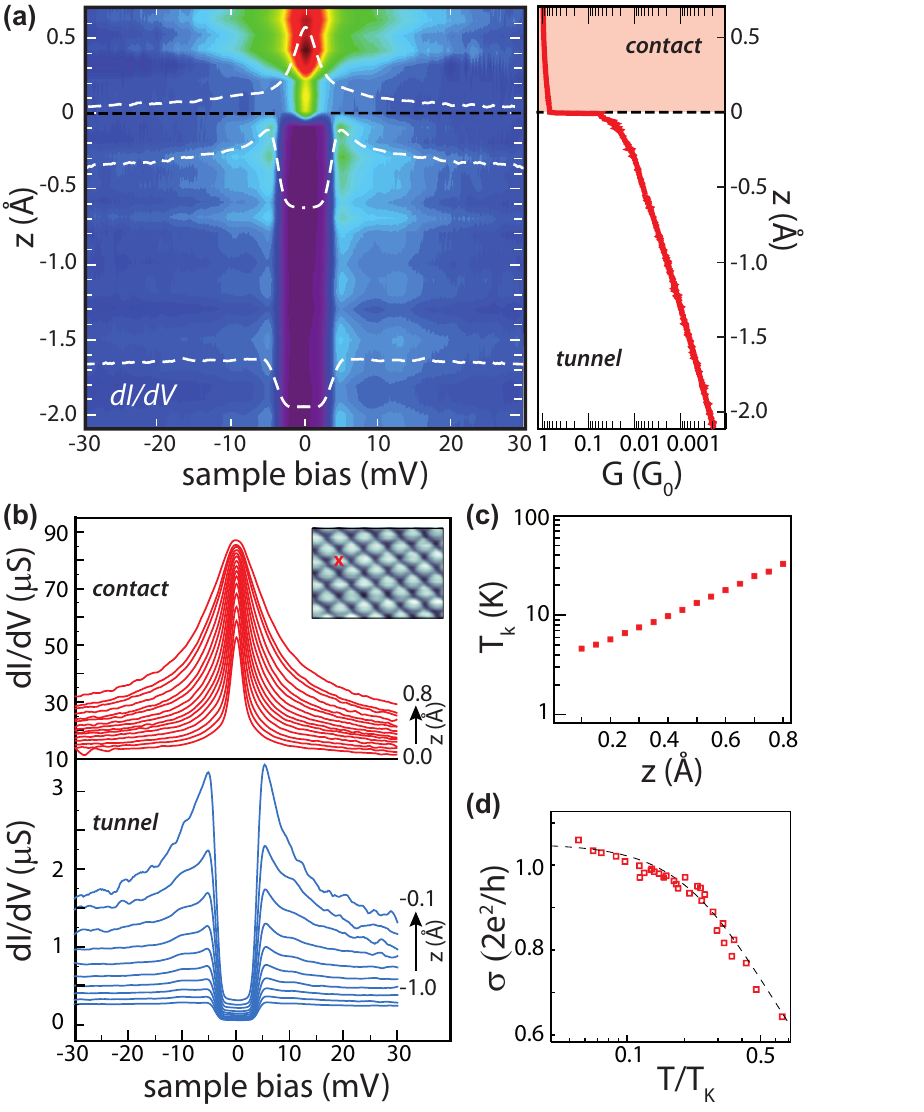}
  \end{center}
  \caption{(a) 2D intensity plot of $dI/dV$ spectra acquired with a Nc-tip at increasing $z$. The intensity of the spectra have been normalized by the opening conductance value (at -30\,mV) of the spectrum at $z$=0\,\AA. As a guideline to identify the differences, three characteristic spectra (white dashed lines) are superimposed. (b) $G$-versus-$z$ curve measured at -2\,mV for a Nc-tip approaching the Cu(100) surface. The boundary between the tunnel and the contact regime occurs at $z$=0\,pm and is indicated with a dashed line. (c) Individual $dI/dV$ spectra in the tunnel (bottom panel) and contact (top panel) regimes for several $z$. Note that right after contact, the width of the Kondo resonance is smaller than the inelastic excitation threshold. Inset: Contact image of the Cu(100) surface using a Nc-terminated tip ($2\times1.4$ nm$^2$; 300\,pA, 30\,mV). The cross indicates the contact point. (d) Evolution of $T_\text{K}$ with $z$. The Kondo temperature is $4.5$~K or lower at $z=0$, and 63.4\,K at $z=1$~\AA. (e) Evolution of the resonance maximum $\sigma$ with $T/T_\text{K}$, where $T=2.4$~K is the working temperature. $\sigma$ and $T_\text{K}$ are extracted from the Frota fits (see Supplementary Material).}
\label{fig2}
\end{figure}

We then engineered a molecular junction by bringing the Nc-tip into contact with Cu(100). For an increased control over the molecular junction, the contact was always performed atop a copper surface atom thanks to the use of atomically-resolved images [Inset of Fig.\,\ref{fig2}\,(b)]. These images were routinely acquired by scanning the molecular tip while in contact with the surface~\cite{Schull2011,Zhang2011}, which may be taken as an indication of the robustness of a Nc-tip to external strain. The right panel of Fig.\,\ref{fig2}\,(a) presents the conductance ($G$) versus tip displacement ($z$) curve acquired at a fixed bias of $-2$\,mV with a Nc-tip vertically displaced towards the surface. The tip is here moved from its initial tunneling position $z=-2.1$\,\AA ($G=2\times10^{-4}$ in units $2e^2/h$) up to $z=0$ ($G=0.04$) where an abrupt increase of $G$ by more than a factor $10$ ($G=0.7$) reveals the transition between the tunneling and the contact regime (indicated by a dashed line). Notice that the exact values of the conductance as well as of the contact position under the same conditions present some tip dependancy. As we show below, the sudden change in $G$ is exclusively driven by a spin switch of nickelocene.

To elucidate this finding, we present in the left panel of Fig.\,\ref{fig2}\,(a) a two-dimensionnal intensity plot of a series of $dI/dV$ spectra acquired at varying $z$. For completeness, Fig.\,\ref{fig2}\,(b) presents the spectra for decreasing tip-surface distances. The spectra reveal the presence of spin-flip scattering events, whose nature may be reversibly controlled via $z$. Within the tunneling regime, the spectra exhibit inelastic spin-flip excitations as anticipated in Fig.~\ref{fig1}(d). We note an enhancement of the differential conductance at voltages corresponding to the excitation threshold, which is increasingly pronounced as the contact point is approached ($z>-0.5$~\AA). Recent studies have pointed out that this enhancement could reflect many-body effects including Kondo-like phenomena~\cite{Hurley2011,Korytar2012,Ternes2015}. However, this scenario also requires a sizable reduction of the magnetic anisotropy energy, and therefore of the excitation threshold~\cite{Oberg2014,Jacobson2015}, at variance with our experimental observations. The enhancement here rather reflects a stationary population of excited states~\cite{Loth2010,Delgado2010,Novaes2010,Ternes2015} due to the increasingly large currents flowing through the junction when approaching contact. This so-called spin pumping can be expected to be significant due to the efficiency of the spin-flip excitations in Nc produced by tunneling electrons~\cite{Ormaza2017}, or more generally, in molecules~\cite{Chen2008,Tsukahara2009,Gauyacq2010}.

\begin{figure}[t]
  \begin{center}
    \includegraphics[width=0.47\textwidth]{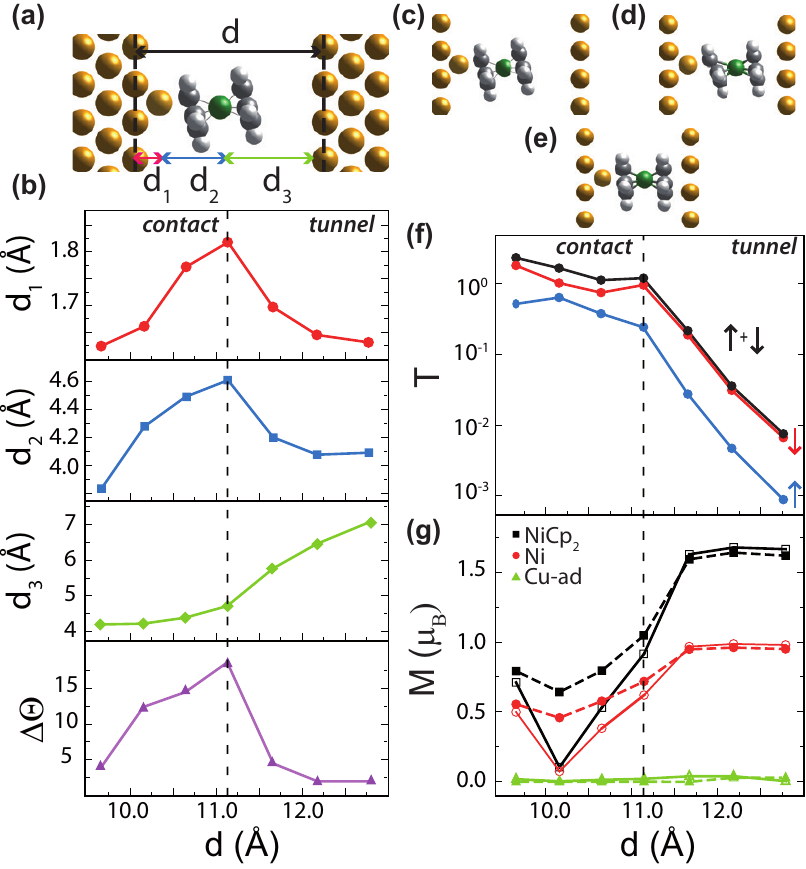}
  \end{center}
  \caption{(a) Structure of the calculated molecular junction, including the main relevant distance parameters. (b) The panels show how the structural parameters $d_1$, $d_2$, $d_3$, $\Delta\Theta$ change as a function of $d$. The structure of the molecular junction for: (c) the tunneling regime ($d$=$12.17$\,\AA), (d) the transition between the tunneling and the contact regimes ($d$=11.14\,\AA), and (e) the contact regime ($d$=9.66\,\AA). (f) Total transmission (black line) and the transmission per spin channel (red and blue lines) at the Fermi level for the explored configurations. (g) Magnetization of: Nc (black), Ni (red), Cu atom (green) as a function of $d$. Full symbols correspond to SIESTA (Mulliken) results and open symbols to VASP (Bader) results.}
\label{fig3}
\end{figure}

The most striking result is found once the molecular contact to the surface is established ($z>0$). As we anticipated above, a sudden increase in the conductance occurs at zero bias [Fig.\,\ref{fig2}\,(a)], which we ascribe to the presence of a Kondo resonance. As shown in Fig.\,\ref{fig2}\,(a), a sharp peak appears in the $dI/dV$, while the inelastic excitation thresholds are lost. The emergence of a spin-1 Kondo effect in the presence or in the absence of positive magnetic anisotropy is in principle possible~\cite{Zitko2008,Misiorny2012} \textemdash the former case being unlikely here as the resonance does not split apart~\cite{Parks2010}. However, we find that the system actually behaves as a spin $1/2$-system. The peak is perfectly fitted by a Frota function (see Supplementary Material), which is close to the exact line shape expected for a spin-$1/2$ Kondo system ~\cite{frota92}. From the fit, we find that the peak is centered at $\epsilon_\text{K}=(0.0\pm0.1)$~meV for all studied tips up to the highest tip excursion explored (0.8\,\AA). The line width, therefore $T_\text{K}$, increases nearly exponentially with $z$ [Fig.\,\ref{fig2}(c)], similarly to other junctions comprising a single Kondo impurity~\cite{DJ2012,DJ2016,DJ2017}. To further confirm the spin-1/2 nature of the Kondo effect, we recall that the resonance amplitude (noted $\sigma$) should be a universal function of the normalized temperature $T/T_\text{K}$. For a quantitative analysis we therefore fit the curve in Fig.\,\ref{fig2}\,(d) to the function $\sigma=[1+(T/T_\text{K})^2 (2^{1/s}-1) ]^{-s}$ (in units of $2e^2/h$)~\cite{Gordon1998} and extract $s=(0.29\pm0.02)$, in remarkable agreement with the spin-1/2 Kondo effect of semiconductor quantum dots~\cite{Wiel2000}. The amplitude of the resonance is close to the unitary limit and indicates a complete Kondo screening~\cite{n2}. Our findings therefore strongly support the emergence of a spin-1/2 Kondo effect where there is no magnetic anisotropy. To elucidate its origin, here below, we show through DFT calculations that Nc switches its spin from $1$ to $1/2$ upon contact with the surface.

The DFT calculations were performed by modeling the molecular tip by a Nc atop a Cu atom adsorbed on a Cu(100) plane [Fig.\,\ref{fig3}\,(a)]. The molecular tip was placed at different distances from a Cu(100) surface and the junction was fully relaxed. Figures\,\ref{fig3}\,(c)-(e) show the resulting molecular junctions for the three most representative configurations. In Fig\,\ref{fig3}\,(c), the distance between the two Cu atomic planes (noted $d$) is $12.17$~{\AA} and corresponds to the tunneling regime [see Fig.~\ref{fig1}(c)] described previously. Figure\,\ref{fig3}\,(d) shows the molecular junction at $d=11.14$~{\AA} and corresponds to the transition between the tunneling and contact regimes. The transition point has been assigned to the point in which a change in the behavior of structural parameters is observed, and has been confirmed by the theoretical transmission probabilities [Fig.\,\ref{fig3}\,(f)]. The molecule is distorted at the transition, but still bonded to the Cu atom through two C atoms. Finally, Fig.\,\ref{fig3}\,(e) corresponds to $d=9.66$~{\AA} and is representative of the molecular junction in the contact regime. The molecule exhibits Cp rings that are parallel to both metallic planes and the Cu atom is coordinated to five C atoms. Figure\,\ref{fig3}\,(b) quantifies the structural changes with $d$. As shown, at the transition point the distances between the Cu and Ni atom with respect to the tip electrode, noted $d_1$ and $d_1+d_2$ respectively, is maximum, as well as the tilt difference between the Cp rings, noted $\Delta\Theta$. The distance between the Ni atom and the surface-electrode (noted $d_3$) instead decreases continuously with $d$. Experimentally, we have seen that in the tunneling regime despite the slightly different initial orientations that the molecule might have on the tip, similar results are obtained in the $dI/dV$ spectra for different molecular tips when going to contact. This indicates that the Nc molecule tends to adopt always the same configuration, with parallel Cp rings, when contacted between the tip and the surface. 

\begin{figure}
  \begin{center}
    \includegraphics[width=0.47\textwidth]{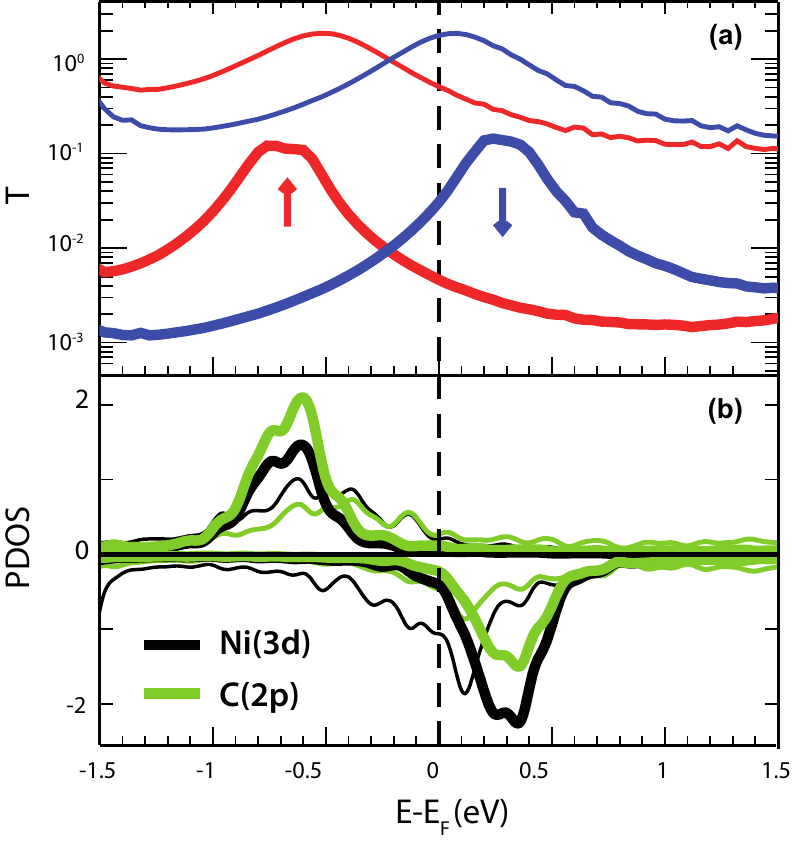}
  \end{center}
  \caption{(a) Spin-resolved electron transmission as a function of electron energy with respect to the Fermi energy for the Nc-tip above Cu(100). The thick line corresponds to the tunneling configuration [Fig.\ref{fig3}\,(f)] and the thin line to the contact configuration [Fig.\ref{fig3}\,(h)]. (b) Density of states projected (PDOS) onto the C(2p) and Ni(3d) atomic orbitals for the tunneling (thick line) and fot the contact (thin line) regimes.}
\label{fig4}
\end{figure}

From Fig.\,\ref{fig3}\,(f) we can obtain the ratio between the transmission probabilities for the different spin channels (spin-up $T_{\uparrow}$, spin-down $T_{\downarrow}$), which is related to the spin polarization efficiency of the molecular junction. We find a -74$\%$ spin polarization for the junction in the tunneling regime and -55$\%$ in contact, where the spin polarization is $\frac{T_{\uparrow}-T_{\downarrow}}{T_{\downarrow}+T_{\uparrow}}$. The decreased value at contact is in agreement with the trend found in a previous theoretical study~\cite{Yi2010}.

Figure\,\ref{fig3}\,(g) presents how the magnetization of the Nc molecule, the Ni atom and the Cu atom change with $d$. In the tunneling regime, the calculated magnetic moment for Nc is $1.7\mu_\text{B}$, the Ni atom carrying $1\mu_\text{B}$. At the transition, these values decrease to $1\mu_\text{B}$ and $0.7\mu_\text{B}$, respectively, and afterwards, at contact, stabilize around $0.75\mu_\text{B}$ for Nc and $0.5\mu_\text{B}$ for the Ni atom. The Cu atom remains non-magnetic during the contact process. The initial magnetic moment of the molecule is halved when contacted between the two electrodes, meaning that the molecular spin changes from $S=1$ to $S=1/2$. Such a spin switch is in agreement with the spectroscopic fingerprints highlighted experimentally.

The change on the molecular charge with respect to the gas phase from tunneling ($-0.12e^-$) to contact ($+0.07e^-$)
is not enough to explain such a change on the magnetization. To scrutinize the effect of the deformation of the molecule 
on the molecular magnetization, the magnetization of the isolated Nc molecule in the same geometrical configuration
as it shows in tunnel, at the transition and in contact was computed. No relevant difference was observed for the different configurations, indicating that the magnetization change is neither driven by the molecule-sustrate charge transfer nor the deformation suffered by the molecule. 

Fig.\ref{fig4}\,(a) presents the spin-resolved transmission for the tunnel (thick line) and contact (thin line) regimes. The transmission of Fig.\,\ref{fig4}\,(a) implies that the transport is mainly due  to the hybridization of surface electronic states with the frontier molecular orbitals, which are $d_{xz}-$ and $d_{yz}-$based molecular orbitals~\cite{Ormaza2017}. To get a deeper understanding of the different features observed in the transmission function, the density of states projected (PDOS) onto the C($2p$) and Ni($3d$) atomic orbitals are shown in Fig.\,\ref{fig4}\,(b). The clear correspondence between spin up and spin down peaks on the transmission function and the PDOS [Fig.\,\ref{fig4}\,(b)] makes it possible to assign both peaks to the transmission through  the spin-polarized degenerate frontier orbitals. The comparison between contact and tunnel results shows that the peaks associated to the frontier-orbital in both spin up and spin down channels shift towards the Fermi level in the contact regime, reducing the exchange splitting energy due to the screening of the intra-orbital Coulomb repulsion ($U$).  This is due to the enhancement of the molecule-electrode interaction that also induces a more pronounced broadening of the molecular levels. Such increase of the hybridization ($\Gamma$) reduces the charge in the majority spin and increases the minority spin occupation. Our analysis  then shows that both effects, namely, the increase of $\Gamma$ and the reduction of $U$ contribute to halve the magnetic moment of the molecule in the contact regime. Therefore, the spin transition can be explained by the coupling to the substrate of the frontier orbitals located close to the Fermi level.


In summary, we have shown how the spin of a Nc molecule and, associated, spin-flip scattering from Nc can be reversibly modified via a controlled contact to a copper electrode. The spin transition takes place due to the enhanced electronic screening following the Nc-surface contact. On a more general note, the portability of the Nc molecule or eventually of other metallocenes~\cite{Ormaza2016} via the STM tip offers novel opportunities for testing how surface-supported objects modify the molecular magnetism of these molecules.



\begin{acknowledgments}
This work has been supported by the Agence Nationale de la Recherche (Grant No. ANR-13-BS10-0016, ANR-11-LABX-0058 NIE, ANR-10-LABX-0026 CSC) the ANPCyT project PICT Bicentenario No. 1962, the CONICET project PIP 0667, and the UNR project PID ING235. We acknowledge computer time provided by
the CCT-Rosario Computational Center and 
the Computational Simulation Center (CSC)
for Technological Applications, members of
the High Performance  Computing National System
(SNCAD, MincyT-Argentina)​. Financial support from MINECO MAT2015-66888-C3-2-R  and FEDER funds is also gratefully acknowledged.
\end{acknowledgments}


\begin{thebibliography}{52}
\bibitem [{n1()}]{n1}%
  \BibitemOpen
  \href@noop {} {}\bibinfo {note} {Note that in a previous work (reference 14)
  in order to determine the magnetic anisotropy of a Nc on top of a Cu atom,
  the molecule was forced to be centered with respect to the Cu adatom
  underneath. However, the current results show that the relaxed situation is
  the one in which the molecule is tilted.}\BibitemShut {Stop}%
\bibitem [{n2()}]{n2}%
  \BibitemOpen
  \href@noop {} {}\bibinfo {note} {Note that in a previous work (reference 47)
  non-equilibrium effects were shown to be absent of the Kondo conductance even
  at contact.}\BibitemShut {Stop}%
\end{thebibliography}
%

\end{document}